\documentclass[aps,prl,showpacs,twocolumn,floats]{revtex4}
\usepackage{amssymb}

\usepackage{dcolumn}
\usepackage{bm}

\begin{document}

\bibliographystyle{prsty}

\title{
Theory of spin Hall effect \vspace{-1mm} }
\author{Eugene M. Chudnovsky}
 \affiliation{\mbox{Department of Physics and Astronomy,} \\
\mbox{Lehman College, City University of New York,} \\ \mbox{250
Bedford Park Boulevard West, Bronx, New York 10468-1589, U.S.A.}
\\ {\today}}
\begin{abstract}
An extension of Drude model is proposed that accounts for spin and
spin-orbit interaction of charge carriers. Spin currents appear
due to combined action of the external electric field, crystal
field and scattering of charge carriers. The expression for spin
Hall conductivity is derived for metals and semiconductors that is
independent of the scattering mechanism. In cubic metals, spin
Hall conductivity $\sigma_s$ and charge conductivity $\sigma_c$
are related through $\sigma_s = [2\pi \hbar/(3mc^2)]\sigma_c^2$
with $m$ being the bare electron mass. Theoretically computed
value is in agreement with experiment.
\end{abstract}


\maketitle

It has been a common knowledge in atomic physics that due to
spin-orbit interaction the spatial separation of electrons with
different spin projections can be achieved through scattering of
an unpolarized electron beam by an unpolarized target \cite{Mott}.
Dyakonov and Perel were the first to notice that in the presence
of the electric current the scattering of charge carriers by
impurities in a semiconductor must lead to a similar effect
\cite{DP}. It was subsequently called the spin Hall effect
\cite{Hirsch} and observed in semiconductors
\cite{Kato,Wunderlich} and metals \cite{Tinkham}. A number of
microscopic models have been developed that explain spatial
separation of spin polarizations by various ``extrinsic'' (due to
impurities) and ``intrinsic'' (impurity-free) mechanisms, see for
review Ref. \onlinecite{Rashba}. While these models provide
valuable insight into microscopic origin of the spin Hall effect,
they are lacking universality of, e.g., Drude model of charge
conductivity \cite{Drude}. The Drude model, in spite of being
classical in nature, has been very powerful in describing dc and
ac conductivity and its temperature dependence. It also gives the
accurate value of the Hall coefficient by catching correctly the
orbital motion of charge carriers in the presence of the magnetic
field. The power of the Drude model resides in the fact that it
expresses conductivity, $\sigma_D = e^2n\tau/m$, via charge $e$,
concentration $n$, mass $m$, and relaxation time $\tau$ of charge
carriers regardless of the scattering mechanism. Same parameters
enter expressions describing experiments other than the Ohm's law,
e.g., $n$ and the sign of $e$ can be extracted from measurements
of the Hall coefficient $R_H=-(nec)^{-1}$, $\tau$ can be extracted
from measurements of the frequency dependence of the impedance,
and $m$ can be extracted from measurements of the cyclotron
resonance. This allows one to test theoretical concepts of charge
conductivity regardless of the degree of accuracy with which one
can compute parameters entering $\sigma_D$.

In this Letter we will try to develop a similar approach to the
spin Hall conductivity. We will take the Drude model a little
further by incorporating spin and spin-orbit interaction into the
dynamics of charge carriers. We will argue that such a
straightforward extension of the Drude model allows one to obtain
universal expression for spin Hall conductivity that is
independent of the scattering mechanism. The spin Hall effect
appears due to combined action of the external electric field,
quadrupole crystal electric field and scattering of charge
carriers. Same as for charge conductivity, all details of the
scattering mechanism are absorbed into the momentum relaxation
time $\tau$. We will show that this crude model provides correct
values of spin Hall conductivity in both metals and
semiconductors.

For certainty we will speak about electrons but the model will
equally apply to holes. The non-relativistic limit of the Dirac
Hamiltonian for spin-1/2 particle is \cite{Drell}
\begin{equation}\label{ham}
{\cal{H}} = \frac{{\bf p}^2}{2m} + U({\bf r}) +
\frac{\hbar}{4m^2c^2}{\bm \sigma}\cdot\left[\frac{\partial
U}{\partial {\bf r}} \times {\bf p}\right]\,.
\end{equation}
Here $U({\bf r})$ represents the action of microscopic electric
field on charge carriers. It incorporates effects of electrostatic
crystal potential $\Phi_0$, potential due to imperfections of the
crystal lattice $\Phi_i$, and external potential $\Phi_e$,
\begin{equation}
U({\bf r}) = e\Phi_0({\bf r}) + e\Phi_i({\bf r}) + e\Phi_e({\bf
r})\,.
\end{equation}
The last term in Eq.\ (\ref{ham}) is the spin-orbit interaction,
with ${\bm \sigma}$ being Pauli matrices.

Hamiltonian mechanics for canonically conjugated variables ${\bf
p}$ and ${\bf r}$ is described by the equations
\begin{eqnarray}
\dot{\bf r} & = & \frac{\partial {\cal{H}}}{\partial {\bf p}} =
\frac{\bf p}{m} + \frac{\hbar}{4m^2c^2}\left[{\bm
\sigma}\times \frac{\partial U}{\partial {\bf r}}\right] \label{r-motion}\\
\dot{\bf p} & = & -\frac{\partial {\cal{H}}}{\partial {\bf r}} =
-\frac{\partial U}{\partial{\bf r}} -
\frac{\hbar}{4m^2c^2}\frac{\partial}{\partial {\bf
r}}\left(\left[{\bm \sigma}\times \frac{\partial U}{\partial {\bf
r}}\right]\cdot{\bf p}\right)\label{p-motion}
\end{eqnarray}
(These equations can also be derived from quantum-mechanical
relations: $i\hbar\dot{\bf r} = [{\bf r},{\cal{H}}]$,
$i\hbar\dot{\bf p} = [{\bf p},{\cal{H}}]$, $[r_i,p_j] =
i\hbar\delta_{ij}$.) From Eq.\ (\ref{r-motion}) one has
\begin{eqnarray}
{\bf p} & = & m\dot{\bf r} - \frac{\hbar}{4mc^2}\left[{\bm
\sigma}\times \frac{\partial U}{\partial {\bf r}}\right] \label{p}
\\
\dot{\bf p} & = & m\ddot{\bf r} - \frac{\hbar}{4mc^2}
\left(\dot{\bf r}\cdot\frac{\partial}{\partial {\bf
r}}\right)\left[{\bm \sigma}\times \frac{\partial U}{\partial {\bf
r}}\right] \label{pdot}
\end{eqnarray}
The second term in the expression for the momentum, Eq.\
(\ref{p}), that is proportional to the cross product of the
electron magnetic moment and the electric field, is the so-called
``hidden mechanical momentum'' associated with the momentum of the
electromagnetic field, see e.g. Ref. \onlinecite{Jackson} and
references therein.

Substitution of Eqs.\ (\ref{p}) and (\ref{pdot}) into Eq.\
(\ref{p-motion}) yields the following form of the second Newton's
law for charge carriers:
\begin{equation}\label{Newton-0}
m\ddot{\bf r} = -\frac{\partial U}{\partial{\bf r}}
 +  {\bf F}_{\sigma}({\bf r}, \dot{\bf
r})\,,
\end{equation}
where the spin-dependent force is given by
\begin{eqnarray}\label{spin-force}
&& {\bf F}_{\sigma}({\bf r}, \dot{\bf r}) = \frac{\hbar}{4mc^2} \times \nonumber \\
&& \left\{\left(\dot{\bf r}\cdot\frac{\partial}{\partial {\bf
r}}\right)\left[{\bm \sigma}\times \frac{\partial U}{\partial {\bf
r}}\right] - \frac{\partial}{\partial {\bf r}}\left(\dot{\bf
r}\cdot\left[{\bm \sigma}\times \frac{\partial U}{\partial {\bf
r}}\right]\right)\right\} \nonumber \\
&& = - \frac{\hbar}{4mc^2}\,\dot{\bf r} \times
\left[\frac{\partial}{\partial {\bf r}} \times \left({\bm
\sigma}\times \frac{\partial U}{\partial {\bf r}}\right)\right]
\,.
\end{eqnarray}
Here we neglected the term proportional to $1/c^4$ that exceeds
the accuracy of Eq.\ (\ref{ham}) and took into account that
$\dot{\bf r}$ in the second line of Eq.\ (\ref{spin-force}), that
originates from the expression for ${\bf p}$ in Eq.\ (\ref{p}),
should not be differentiated on ${\bf r}$ because ${\bf p}$ and
${\bf r}$ in Eqs.\ (\ref{r-motion}) and (\ref{p-motion}) are
independent canonically conjugated variables. Note that the force
in Eq.\ (\ref{spin-force}) is equivalent to the Lorentz force,
${\bf F}_{\sigma} = (e/c)(\dot{\bf r}\times{\bf B}_{\sigma})$,
acting on a particle of charge $e$ in the magnetic field
\begin{equation}
{\bf B}_{\sigma} = {\bm \nabla} \times{\bf A}_{\sigma}\,, \qquad
{\bf A}_{\sigma} = \frac{\hbar}{4mc}\left({\bm \sigma}\times {\bf
E}_{tot}\right)\,,
\end{equation}
with ${\bf E}_{tot}$ being the total electric field, $e{\bf
E}_{tot} = -{\partial}U/{\partial}{\bf r}$. One can trace this
force to the fact that with an accuracy to $c^{-2}$ the
Hamiltonian (\ref{ham}) can be written as \cite{AC}
\begin{equation}
{\cal{H}} = \frac{1}{2m}\left({\bf p} - \frac{e}{c}{\bf
A}_{\sigma}\right)^2 + U({\bf r})\,.
\end{equation}
Crystal field creates a non-zero average of this force any time
the charge carriers have a non-zero drift velocity,
$\langle\dot{\bf r}\rangle \neq 0$. {\it It is this effective
Lorentz force that is responsible for the spin Hall effect.} This
our conclusion is similar to the conclusion of Hirsch
\cite{Hirsch-99} who studied the force exerted on a line of moving
magnetic dipoles by the electrostatic field of charges arranged in
a cubic lattice. For a cubic lattice (see below) our result for
the effective Lorentz force coincides up to a factor of 2 with the
result obtained by Hirsch. As in our approach, Hirsch found that
the effective Lorentz force is generated by the second derivative
of the crystal field. In computing this force he replaced moving
magnetic dipoles with stationary electric dipoles that produce an
equivalent electric field. This eliminated the hidden momentum
responsible for the first term in the second line of Eq.\
(\ref{spin-force}). The absence of this term in Hirsch's model
accounts for the above mentioned difference by a factor of 2.

In the spirit of the Drude model we shall now add to Eq.\
(\ref{Newton-0}) the drag force $-m\dot{\bf r}/\tau$. Then Eq.\
(\ref{Newton-0}) becomes
\begin{equation}\label{Newton}
m\ddot{\bf r} = -\frac{\partial U}{\partial{\bf r}}
-\frac{m}{\tau}\dot{\bf r} +  {\bf F}_{\sigma}({\bf r}, \dot{\bf
r})\,.
\end{equation}
Here we assume that to the first approximation the velocity
relaxation of charge carriers is independent of their spin, that
is, $\tau$ is independent of ${\bm \sigma}$. Since the relaxation
is due to imperfections of the crystal lattice, in order not to
count their effect twice, we should now think of $U({\bf r})$ in
Eqs.\ (\ref{Newton}) and (\ref{spin-force}) as a sum of the ideal
periodic potential of the crystal lattice, $e\Phi_0$, and the
potential produced by the externally applied voltage, $e\Phi_e$.
Due to relativistic smallness of the spin-dependent force
(\ref{spin-force})  one can treat ${\bf F}_{\sigma}({\bf r},
\dot{\bf r})$ in Eq.\ (\ref{Newton}) as a perturbation. Then the
solution of Eq.\ (\ref{Newton}) can be written in the form
$\dot{\bf r} = \dot{\bf r}_0 + \dot{\bf r}_1$, where $\dot{\bf
r}_1$ is a small spin-dependent part of the velocity proportional
to $c^{-2}$. In the presence of a constant external electric field
${\bf E} = -\partial \Phi_e/\partial {\bf r}$, with the linear
accuracy on ${\bf E}$, one obtains from Eqs.\ (\ref{Newton}) and
(\ref{spin-force})
\begin{eqnarray}
\langle \dot{\bf r}_0 \rangle  & = & - \frac{\tau}{m}\left\langle
\frac{\partial U}{\partial{\bf r}}\right\rangle = \frac{e\tau}{m}
{\bf
E} \label{r0} \\
\langle \dot{\bf r}_1 \rangle & = & \frac{\tau}{m}\left\langle{\bf
F}_{\sigma}({\bf r}_0, \dot{\bf r}_0)\right\rangle\nonumber \\
& = & -\frac{\hbar e^2 \tau^2}{4m^3c^2}\,{\bf E}\times\left\langle
\frac{\partial}{\partial {\bf r}}\times\left({\bm \sigma}\times
\frac{\partial \Phi_0}{\partial {\bf r}}\right) \right\rangle.
\label{r1}
\end{eqnarray}
In deriving Eq.\ (\ref{r1}) we have made an assumption that
\begin{equation}\label{factorization}
\langle{\bf F}_{\sigma}({\bf r}_0, \dot{\bf r}_0)\rangle \equiv
\frac{e}{c}\left\langle\dot{\bf r}_0\times{\bf B}_{\sigma}({\bf
r}_0) \right\rangle = \frac{e}{c}\left\langle\dot{\bf
r}_0\right\rangle\times\left\langle{\bf B}_{\sigma}
\right\rangle\,.
\end{equation}
Some justification of this assumption is provided by the following
argument. By symmetry, the only reason for $\langle{\bf
F}_{\sigma}\rangle$ to be different from zero would be
$\langle\dot{\bf r}_0\rangle \neq 0$. Consequently, $\langle{\bf
F}_{\sigma}\rangle$ should be first order on $\langle\dot{\bf
r}_0\rangle$. Being perpendicular to the velocity, the force ${\bf
F}_{\sigma} = (e/c)(\dot{\bf r}\times{\bf B}_{\sigma})$ does not
do mechanical work on the charge. Neither should $\langle{\bf
F}_{\sigma}\rangle$ with respect to $\langle\dot{\bf r}_0\rangle$,
rendering the form $\langle{\bf F}_{\sigma}\rangle =
(e/c)(\langle\dot{\bf r}_0\rangle\times{\bf B}_{eff})$. Since the
trajectory of the particle ${\bf r}_0(t)$ does not have strong
correlation with the quadrupole component of the crystal electric
field, the above factorization of the average with the choice
${\bf B}_{eff} = \langle{\bf B}_{\sigma} \rangle$ should not
deviate strongly from the exact average.

The right hand side of Eq.\ (\ref{r1}) contains the volume average
of $\nabla_i\nabla_j\Phi_0({\bf r})$. In what follows we will
study the spin Hall effect in a cubic lattice. Generalization to
other lattices is straightforward and will be considered
elsewhere. The case of a cubic lattice, besides simplicity, is
interesting because experiments performed to date have been done
in cubic semiconductors and in aluminum that is also cubic
\cite{Kato,Wunderlich,Tinkham}. In the cubic case the only
invariant permitted by symmetry is
\begin{equation}\label{Phi-ij}
\left\langle\frac{\partial^2 \Phi}{\partial r_i \partial
r_j}\right\rangle = A \delta_{ij}\,,
\end{equation}
where $A$ is a constant to be determined later. With the help of
Eq.\ (\ref{Phi-ij}) one obtains from Eq.\ (\ref{r1}):
\begin{equation}\label{spin-velocity}
\langle \dot{\bf r}_1 \rangle = \frac{\hbar e^2 \tau^2
A}{2m^3c^2}[{\bm \sigma}\times {\bf E}]\,.
\end{equation}

Let us now introduce the vector of spin polarization of the
electron fluid, ${\bm \xi} = \langle{\bm \sigma}\rangle$, which
absolute value lies between $0$ and $1$,
\begin{equation}\label{polarization}
\xi = \frac{n_+ - n_-}{n_+ + n_-}\,.
\end{equation}
Here $n_{\pm}$ are concentrations of charge carriers with spins
parallel and antiparallel to ${\bm \xi}$ respectively, with $n_+ +
n_- = n$ being the total concentration of charges carrying the
electric current. For a ferromagnet, $\xi \neq 0$ is the
equilibrium state of charge carriers below the Curie temperature,
while for a small non-magnetic conductor a significant value of
$\xi$ can be achieved through injection of spin-polarized charge
carriers from a magnetic conductor, see for review Ref.
\onlinecite{DasSarma}. The density matrix of the charge carriers
in the spin space (normalized to their total concentration $n$)
can be written as
\begin{equation}\label{den-matrix}
N = \frac{1}{2}n(1 + {\bm \xi}\cdot{\bm \sigma})\,.
\end{equation}
The electric current is
\begin{equation}
{\bf j} = e \langle N\dot{\bf r}\rangle = e \langle N(\dot{\bf
r}_0 + \dot{\bf r}_1)\rangle\,.
\end{equation}
Substituting here equations (\ref{r0}) and (\ref{spin-velocity})
one obtains
\begin{equation}\label{current}
{\bf j} = \sigma_c {\bf E} + \sigma_{s}[{\bm \xi} \times {\bf
E}]\,,
\end{equation}
where charge conductivity and spin Hall conductivity are given by
\begin{eqnarray}
& & \sigma_c = \sigma_D = \frac{e^2n\tau}{m} \label{chargecond} \\
& & \sigma_{s} = \frac{\hbar e^3 n \tau^2 A}{2m^3c^2}
\label{spincond}
\end{eqnarray}

An interesting observation is that the ratio
\begin{equation}\label{ratio}
\frac{\sigma_s}{\sigma_c} = \frac{\hbar e  \tau A}{2 m^2 c^2}
\end{equation}
is independent of the concentration of charge carriers. This
result also follows from microscopic models of spin Hall effect
\cite{Engel,Tse}. It explains why this ratio has the same order of
magnitude in metals and semiconductors \cite{Kato,Tinkham} despite
very different concentrations of charge carriers. Note that
according to Eq.\ (\ref{ratio}) the temperature dependence of the
ratio of spin Hall and charge conductivities is determined by the
temperature dependence of the relaxation time $\tau$.
Consequently, in metals one should observe monotonic temperature
dependence of spin Hall conductivity, $\sigma_s(T) \propto
\sigma_c^2(T)$, while in semiconductors $\sigma_s(T) \propto
n(T)\tau^2(T)$ may exhibit maximum on temperature \cite{Stern}.

Eq.\ (\ref{r1}) shows that quantitative analysis of the spin Hall
effect requires knowledge of the volume average of the quadrupole
component of the crystal electric field, $\langle\partial^2
\Phi_0/\partial r_i \partial r_j\rangle$. For a specific crystal
it can be computed by methods of density functional theory. For a
cubic crystal $\langle\partial^2 \Phi_0/\partial r_i
\partial r_j\rangle$ reduces to a single
constant $A$, see Eq.\ (\ref{Phi-ij}). Here we will provide a
simple estimate of this constant. In a metal the electrostatic
potential created by the crystal lattice of positively charged
ions satisfies Maxwell equation
\begin{equation}\label{Phi}
{\nabla}^2 \Phi_0 = -4\pi \rho(x,y,z)\,,
\end{equation}
where $\rho(x,y,z)$ is the local charge density of ions. The
volume average of this equation is
\begin{equation}\label{laplacian-Phi}
\left\langle\frac{\partial^2 \Phi_0}{\partial x^2}\right\rangle +
\left\langle\frac{\partial^2 \Phi_0}{\partial y^2}\right\rangle +
\left\langle\frac{\partial^2 \Phi_0}{\partial z^2}\right\rangle =
3A = -4\pi \langle\rho(x,y,z)\rangle
\end{equation}
Due to the periodicity of the crystal field, it is sufficient to
compute this average over the unit cell of the crystal. This gives
\begin{equation}\label{A}
A = \frac{4\pi}{3} Z e n_0\,,
\end{equation}
where $-Ze$ and $n_0$ are the charge and the concentration of ions
respectively. The sign in $\langle \rho \rangle = -Zen_0$ is
determined by our choice of the negative charge $e$ of the charge
carriers. In the accurate quantum-mechanical calculation of the
parameter $A$ produced by the distribution of charges in a cubic
lattice, $Z$ should come out as a number of order unity when $n_0$
is chosen as the inverse volume of the unit cell. This should be
true for both metals and semiconductors.

Substitution of Eq.\ (\ref{A}) into Eq.\ (\ref{ratio}) gives
\begin{equation}\label{ratio-exact}
\frac{\sigma_s}{\sigma_c} = \frac{2\pi \hbar e^2 }{3 m^2 c^2}Zn_0
\tau\,.
\end{equation}
At $n_0 \sim 10^{22}$cm$^{-3}$ ($10^{23}$cm$^{-3}$) and $\tau \sim
10^{-13}$s ($10^{-14}$s) this ratio is of order $10^{-4}$ which is
in agreement with recent experimental findings in metals and
semiconductors \cite{Kato,Tinkham}. According to our model such a
value of ${\sigma_s}/{\sigma_c}$ is not universal though. In pure
conductors, when scattering of charge carriers is dominated by
phonons, large $\tau$ at low temperature can provide much greater
values of the ratio of spin Hall and charge conductivities.

For the absolute value of spin Hall conductivity one obtains from
Eq.\ (\ref{spincond})
\begin{equation}\label{final}
\sigma_s = \frac{2\pi\hbar e^4}{3m^3c^2}Z n n_0 \tau^2 \,.
\end{equation}
In metals, when $Z$ is identified with the ion valence, one can
replace $Zn_0$ with $n$ and write Eq.\ (\ref{final}) as
\begin{equation}\label{metal}
\sigma_s = \frac{2\pi \hbar}{3mc^2}\sigma_c^2\,.
\end{equation}
This relation between spin Hall and charge conductivities permits
comparison with experimental data without any fitting parameters.
Such a comparison with the data on aluminum \cite{Tinkham} is
shown in Table \ref{table}.
\begin{table}[here]
\begin{tabular}{ccc}
Conductivity  \qquad  & Experiment
($\Omega^{-1}$m$^{-1}$) \qquad & Theory ($\Omega^{-1}$m$^{-1}$)\\
\hline $\sigma_c$ (12nm) \qquad & $1.05 \times 10^7$  \qquad  &
\\ \hline $\sigma_s$ (12nm) \qquad
& $(3.4 \pm 0.6) \times 10^3$ \qquad  & $2.6 \times 10^3$
\\ \hline $\sigma_c$
(25nm) \qquad  & $1.7 \times 10^7$ \qquad  &  \\
\hline $\sigma_s$ (25nm) \qquad  & $(2.7 \pm 0.6) \times 10^3$
\qquad  & $6.9 \times 10^3$
\end{tabular}
\caption{Comparison of Eq.\ (\ref{metal}) with experimental data
\cite{Tinkham} on charge and spin Hall conductivities in Al strips
of $12$-nm and $25$-nm thickness.}\label{table}
\end{table}
(Note that conductivity in Gauss units that have been used to
derive Eq.\ (\ref{metal}), equals conductivity in units of
$\Omega^{-1}$m$^{-1}$ times $9 \times 10^9$.) Given the crudeness
of the model the agreement by order of magnitude between theory
and experiment is quite remarkable. Note that in the spirit of
Drude model one can also make a straightforward extension of our
model to the ac spin Hall conductivity.

Throughout our derivations we did not distinguished between the
effective mass and the bare mass of charge carriers. To account
for their difference we notice that the spin-orbit term in Eq.\
(\ref{ham}) can be interpreted as the Zeeman interaction of the
electron spin with the effective magnetic field that is generated
by the electrostatic field in the moving frame of the electron.
Consequently, one of the masses in $\hbar/(4m^2c^2)$ in front of
the spin-orbit term must be the bare electron mass that enters the
Bohr magneton, while the other $m$ is due to orbital motion. In a
solid, one should replace $m^2$ in the spin-orbit term with
$mm^*$, where $m^*$ is the effective electron mass. Also in the
first term of Eq.\ (\ref{ham}) $m$ should be replaced by $m^*$.
After these replacements are made and followed down to Eq.\
(\ref{metal}), $m^{*2}$ gets absorbed into $\sigma_c^2$ while
$1/m$ in front of $\sigma_c^2$ can be traced to the expression for
the Bohr magneton, $\mu_B = |e| \hbar/(2mc)$. Thus the coefficient
in front of $\sigma_c^2$ in Eq.\ (\ref{metal}) contains the bare
electron mass. We used this fact for computing numerical values of
$\sigma_s$ presented in Table \ref{table}.

The theory of spin Hall effect developed in this Letter should
also apply to the anomalous Hall effect observed in ferromagnets.
According to Eq.\ (\ref{spin-velocity}) the weak flow of charge
carriers with opposite spin polarizations in the opposite
(perpendicular to ${\bf E}$) directions occurs regardless of the
average polarization $\xi$ given by Eq.\ (\ref{polarization}).
When the time of spin relaxation of charge carriers is
sufficiently long, this will lead to the spin polarization of the
boundaries of a small conductor even when $\xi = 0$. On the
contrary, the spin Hall current given by the second term in Eq.\
(\ref{current}) is proportional to $\xi$. In non-magnetic metals
and semiconductors, large spin polarization is not easy to achieve
but in a ferromagnetic metal $\xi$ can be of order unity,
resulting in the noticeable current, $\sigma_s[{\bm \xi}\times
{\bf E}]$, that is normal to ${\bf E}$ and normal to the
magnetization.

In Conclusion, we have presented a simple extension of Drude model
of conductivity to the case of charge carriers that have spin and
spin-orbit interaction. Spin Hall effect appears naturally in such
a theory as a combined action of the external voltage, crystal
electric field, and scattering of charge carriers. The expression
for spin Hall conductivity is independent of the mechanism of
scattering. Theoretical values of spin Hall conductivity computed
within such a model are in agreement with experimental data on
metals and semiconductors. The model can, therefore, serve as a
simple physical picture of spin Hall effect.

The author thanks Dmitry Garanin and Jorge Hirsch for discussion.
This work has been supported by the DOE grant No.
DE-FG02-93ER45487.

\end{document}